\documentclass[twocolumn,showpacs,preprintnumbers,amsmath,amssymb,superscriptaddress]{revtex4}

\usepackage{graphicx}
\usepackage{dcolumn}
\usepackage{bm}
\usepackage{epsfig}
\usepackage{color}

\newcommand{ \rts }{$\sqrt{s_{_{\rm NN}}}$}
\newcommand{\mpt}{$\langle p_{t} \rangle\,$}
\newcommand{\npart}{$N_{part}$}
\newcommand{\nbin}{$N_{bin}$}
\newcommand{\nch}{$N_{ch}$ }
\newcommand{\rcp}{$R_{CP}$ }
\newcommand{\rdau}{$R_{dAu}$ }
\newcommand{\rcpno}{$R_{CP}$}
\newcommand{\rdauno}{$R_{dAu}$}
\newcommand{\scp}{$S_{CP}(\eta,\eta_{CM})$ }
\newcommand{\sdau}{$S_{dAu}(\eta,\eta_{CM})$ }
\begin{document}

\title{Charged particle distributions and nuclear modification at high rapidities in d+Au collisions at $\sqrt{s_{NN}}$ = 200 GeV}

\affiliation{Argonne National Laboratory, Argonne, Illinois 60439}
\affiliation{University of Birmingham, Birmingham, United Kingdom}
\affiliation{Brookhaven National Laboratory, Upton, New York 11973}
\affiliation{California Institute of Technology, Pasadena, California 91125}
\affiliation{University of California, Berkeley, California 94720}
\affiliation{University of California, Davis, California 95616}
\affiliation{University of California, Los Angeles, California 90095}
\affiliation{Carnegie Mellon University, Pittsburgh, Pennsylvania 15213}
\affiliation{University of Illinois, Chicago}
\affiliation{Creighton University, Omaha, Nebraska 68178}
\affiliation{Nuclear Physics Institute AS CR, 250 68 \v{R}e\v{z}/Prague, Czech Republic}
\affiliation{Laboratory for High Energy (JINR), Dubna, Russia}
\affiliation{Particle Physics Laboratory (JINR), Dubna, Russia}
\affiliation{University of Frankfurt, Frankfurt, Germany}
\affiliation{Institute of Physics, Bhubaneswar 751005, India}
\affiliation{Indian Institute of Technology, Mumbai, India}
\affiliation{Indiana University, Bloomington, Indiana 47408}
\affiliation{Institut de Recherches Subatomiques, Strasbourg, France}
\affiliation{University of Jammu, Jammu 180001, India}
\affiliation{Kent State University, Kent, Ohio 44242}
\affiliation{Institute of Modern Physics, Lanzhou, China}
\affiliation{Lawrence Berkeley National Laboratory, Berkeley, California 94720}
\affiliation{Massachusetts Institute of Technology, Cambridge, MA 02139-4307}
\affiliation{Max-Planck-Institut f\"ur Physik, Munich, Germany}
\affiliation{Michigan State University, East Lansing, Michigan 48824}
\affiliation{Moscow Engineering Physics Institute, Moscow Russia}
\affiliation{City College of New York, New York City, New York 10031}
\affiliation{NIKHEF and Utrecht University, Amsterdam, The Netherlands}
\affiliation{Ohio State University, Columbus, Ohio 43210}
\affiliation{Panjab University, Chandigarh 160014, India}
\affiliation{Pennsylvania State University, University Park, Pennsylvania 16802}
\affiliation{Institute of High Energy Physics, Protvino, Russia}
\affiliation{Purdue University, West Lafayette, Indiana 47907}
\affiliation{Pusan National University, Pusan, Republic of Korea}
\affiliation{University of Rajasthan, Jaipur 302004, India}
\affiliation{Rice University, Houston, Texas 77251}
\affiliation{Universidade de Sao Paulo, Sao Paulo, Brazil}
\affiliation{University of Science \& Technology of China, Hefei 230026, China}
\affiliation{Shanghai Institute of Applied Physics, Shanghai 201800, China}
\affiliation{SUBATECH, Nantes, France}
\affiliation{Texas A\&M University, College Station, Texas 77843}
\affiliation{University of Texas, Austin, Texas 78712}
\affiliation{Tsinghua University, Beijing 100084, China}
\affiliation{Valparaiso University, Valparaiso, Indiana 46383}
\affiliation{Variable Energy Cyclotron Centre, Kolkata 700064, India}
\affiliation{Warsaw University of Technology, Warsaw, Poland}
\affiliation{University of Washington, Seattle, Washington 98195}
\affiliation{Wayne State University, Detroit, Michigan 48201}
\affiliation{Institute of Particle Physics, CCNU (HZNU), Wuhan 430079, China}
\affiliation{Yale University, New Haven, Connecticut 06520}
\affiliation{University of Zagreb, Zagreb, HR-10002, Croatia}

\author{B.I.~Abelev}\affiliation{University of Illinois, Chicago}
\author{M.M.~Aggarwal}\affiliation{Panjab University, Chandigarh 160014, India}
\author{Z.~Ahammed}\affiliation{Variable Energy Cyclotron Centre, Kolkata 700064, India}
\author{B.D.~Anderson}\affiliation{Kent State University, Kent, Ohio 44242}
\author{D.~Arkhipkin}\affiliation{Particle Physics Laboratory (JINR), Dubna, Russia}
\author{G.S.~Averichev}\affiliation{Laboratory for High Energy (JINR), Dubna, Russia}
\author{Y.~Bai}\affiliation{NIKHEF and Utrecht University, Amsterdam, The Netherlands}
\author{J.~Balewski}\affiliation{Indiana University, Bloomington, Indiana 47408}
\author{O.~Barannikova}\affiliation{University of Illinois, Chicago}
\author{L.S.~Barnby}\affiliation{University of Birmingham, Birmingham, United Kingdom}
% ...

\collaboration{STAR Collaboration}\noaffiliation

\begin{abstract}

The measurements of the centrality dependence of $dN/d\eta$ and transverse momentum spectra from mid- to forward rapidity in d+Au collisions at \rts $=$ 200 GeV are reported. They provide a sensitive tool for understanding the dynamics of multi-particle production in the high parton-density regime. In particular, we observe strong suppression of the nuclear modification factor \rcp at forward rapidities (d-side, $\eta$ = 3.1) and enhancement at backward rapidity ($\eta$ = $-$3.1). An empirical scaling is obtained for multiplicity and \rcp when a shift of the center-of-mass in the asymmetric d+Au collisions with respect to the nucleon-nucleon system is applied.

\end{abstract}

\maketitle


\section{\label{intro}Introduction}
The d+Au collisions at the Relativistic Heavy Ion Collider (RHIC) provide an important control environment compared to Au+Au collisions. Measurements of the nuclear modification factor and back-to-back correlations at mid-rapidity in d+Au collisions suggest that the suppression of particles with high transverse momentum and the disappearance of back-to-back correlations, seen in Au+Au collisions \cite{star_jet_auau}, are due to final-state interactions with the hot, dense medium produced in such collisions, rather than initial-state effects on the Au nucleus \cite{saturation_auau_dau}. The observed enhancement of the nuclear modification factor in the region of transverse momentum ($p_{t}$) of 2 GeV/c at mid-rapidity in d+Au collisions \cite{star_dAu,phob_dAu,phenix,brahms_dAu_highpt}, referred to as the ``Cronin effect'' \cite{cronin}, can be described within a pQCD framework incorporating initial multiple parton scattering and nuclear shadowing \cite{vitev_dau,wang_dAu}. Saturation effects (mostly described as the formation of Color Glass Condensate (CGC) \cite{saturation_auau}) are expected to be more pronounced at large rapidity ($y$) or pseudorapidity ($\eta$) close to the deuteron beam, where the small-$x$ components of the Au nucleus wave function can be probed. Recent results reported by the BRAHMS collaboration \cite{brahms_dAu_rcp}, where a suppression of the nuclear modification factor at forward rapidities is visible, are in qualitative agreement with predictions within the framework of gluon saturation in the CGC \cite{saturation_dAu}.  These results indicate a possible dramatic evolution of gluon saturation from mid- to forward rapidity at RHIC. However, it should be noted that these results can be reasonably described by pQCD models \cite{vitev_dau,wang_dAu} and in the framework of final-state parton recombination \cite{recombination}.  On the other hand, as the rapidity of the probe decreases (and at the same time, $x$ increases), the multiple-scattering contribution to the Cronin effect should decrease.  The first results from the PHENIX collaboration \cite{phenix_rcp} show an opposite behaviour which cannot be explained by current model calculations \cite{accardi04}. 

In this paper the pseudorapidity and centrality dependence of the nuclear modification factor \rcp will be discussed in connection with the asymmetry in particle production in d+Au collisions at $\sqrt{s_{NN}}$ =200 GeV.

\section{\label{exp}Experimental Setup}

The STAR experiment \cite{Ackermann2003} at RHIC measures charged hadrons over a wide range of pseudorapidity and transverse momentum. The main detector is a large Time Projection Chamber (TPC) \cite{Anderson2003a} which allows particle identification via dE/dx within the range of $|\eta|<$ 1. Charged particle detection in the forward directions is achieved with the two azimuthally symmetric Forward TPC (FTPCs \cite{Ackermann2003a}) which extend the pseudorapidity coverage of STAR to the region 2.5 $<|\eta|<$ 4. The FTPCs which utilize a radial drift field perpendicular to the magnetic field, achieve a two-track resolution of 2 -- 2.5 mm (an order of magnitude better than a TPC using a constant drift field). This allows track reconstruction in the larger rapidity region where track densities are high. In d+Au collisions at a center-of-mass energy of $\sqrt{s_{NN}}$=200 GeV a track finding efficiency in the FTPCs of about 90\% was reached, independent of centrality, $\eta$ and $p_{t}$ in the phase space region of 3 $< \eta< $ 3.5 and 0.1  $<p_{t}<$ 3 GeV/c used for this analysis. The momentum resolution in the FTPCs is also independent of centrality, but shows a strong dependence on $\eta$ and $p_{t}$. For $\eta \approx$  3.1, the relative momentum resolution degrades approximately linearly from 10\% to 25\% -- 30\% in the region of 0.1  $<p_{t}<$ 3 GeV/c and the pseudorapidity resolution is better than 0.02 units in $\eta$. Background and secondary decay products corrections were estimated using HIJING simulations \cite{hijing}. The main systematic error quoted in this analysis (if not otherwise mentioned) is caused by the momentum resolution of the FTPCs. This affected mainly the vertex DCA (distance of closest approach) requirement used to select primary charged hadrons in the FTPCs. An estimate of the main systematic error was done by varying the DCA by $\pm$ 0.5 cm. A detailed description of the various calibration steps, further corrections and data quality can be found in \cite{thesis}.


\begin{figure*}[t]
\begin{minipage}[t]{18pc}
\includegraphics[width=20pc]{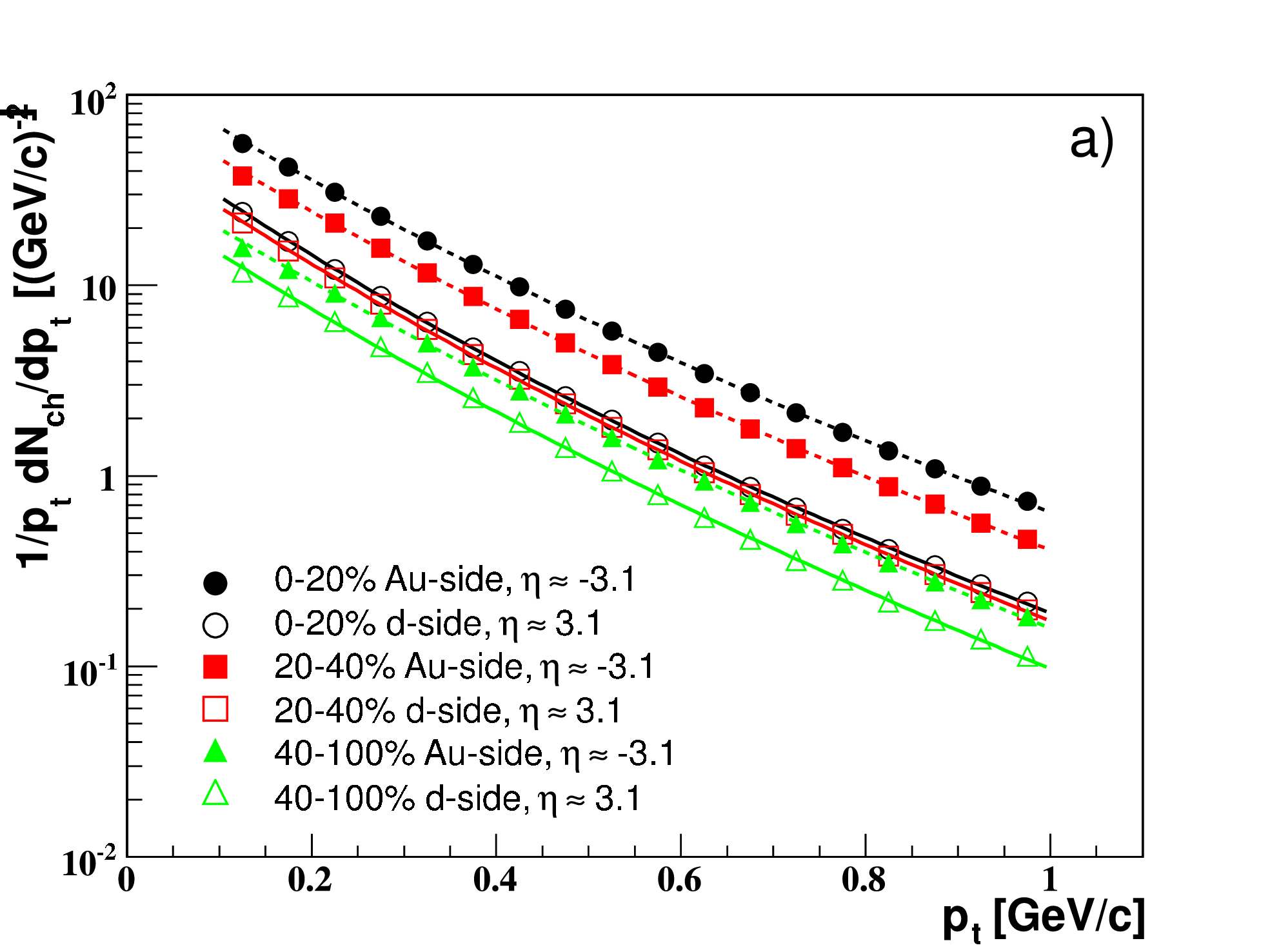}
\end{minipage}\hspace{2pc}%
\begin{minipage}[t]{18pc}
\includegraphics[width=20pc]{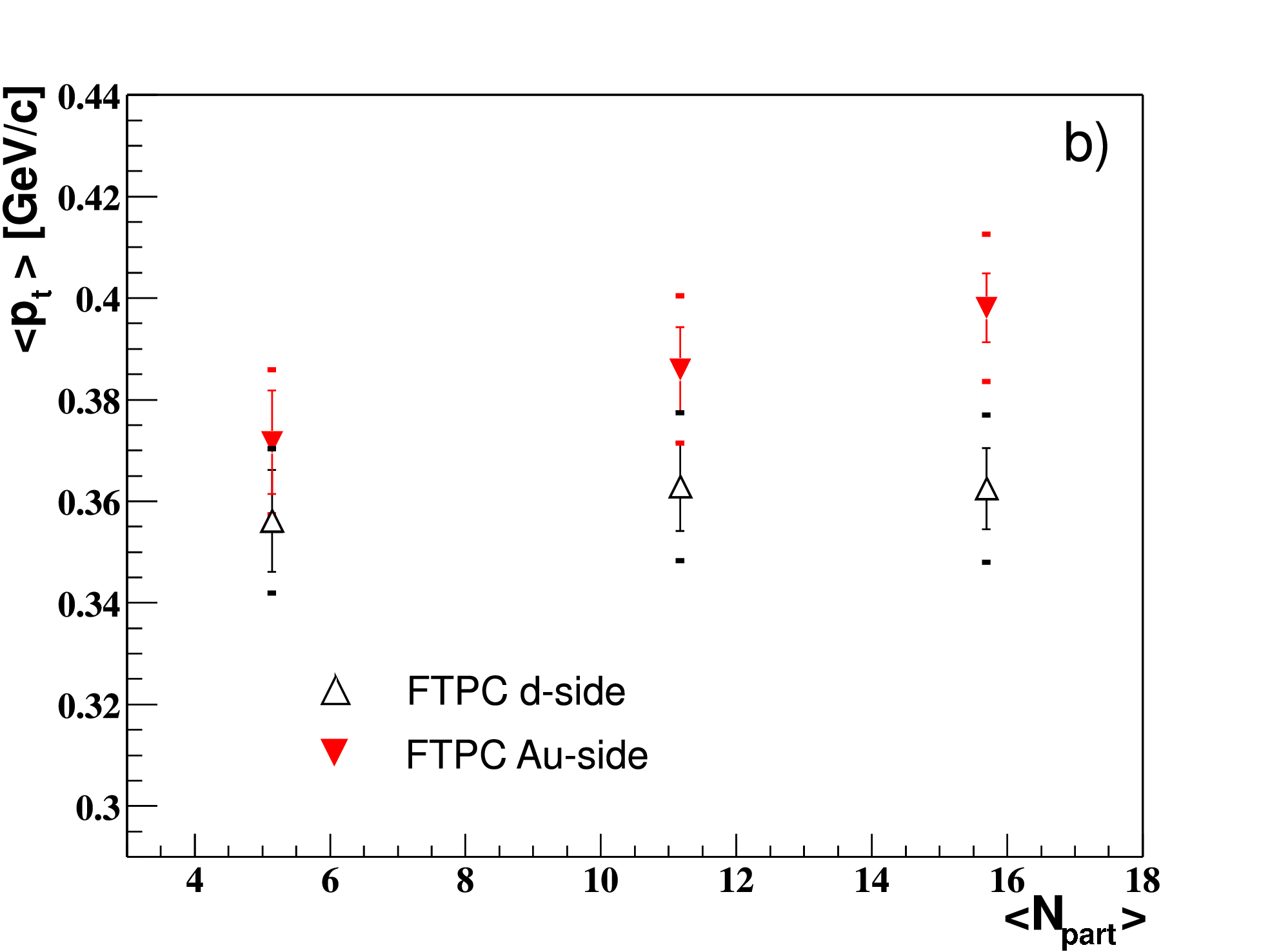}
\end{minipage} 
\caption{\label{fig1} (Color online) a) Charged hadron transverse momentum distributions at $|\eta| \approx$ 3.1 for 0-20\%, 20-40\% and 40-100\% central d+Au events. Closed triangles refer to measurements on the Au-side ($\eta \approx$  $-$3.1) and open triangles to the d-side ($\eta \approx$  3.1). b) \mpt extracted from fitting the transverse momentum spectra with a power law function at $|\eta| \approx$ 3.1 as a function of \npart. The closed triangles are measurements on the Au-side and the open triangles on the d-side of a d+Au collision. Statistical (lines) and systematical (bars) are shown separately (for details see text).}
\end{figure*}

\section{\label{mes}Measurements}

\subsection{\label{mptsec}Mean transverse momentum}

To quantify the influence of momentum resolution on the transverse momentum spectra simulated charged pions were embedded in real d+Au events, adding up to 5\% of the total event multiplicity in the FTPCs. Initially flat input distributions were weighted according to the measured transverse momentum distributions as function of centrality and transverse momentum \cite{thesis}. Background and secondary decay products corrections were estimated using HIJING simulations \cite{hijing,thesis}. The transverse momentum distributions corrected for momentum resolution, background and secondary decay products measured in the East-FTPC (Au-side; $\eta \approx$  $-$3.1) and West-FTPC (d-side; $\eta \approx$  3.1) are shown in Fig.\ \ref{fig1}a for different centrality classes. Fig.\ \ref{fig1}b shows the mean transverse momentum \mpt obtained from a power-law fit to the measured $p_t$ spectra in the region of 0.1 $< p_{t}<$ 1 GeV/c as a function of the number $N_{part}$ of participant nucleons. In contrast to the ``naive" picture in which the partons from the d-nucleus experience multiple collisions while traversing the Au-nucleus, and therefore acquire an enhancement of \mpt on the d-side of the collision \cite{accardi04}, a slight increase of \mpt with centrality is visible on the Au-side, whereas on the d-side virtually no centrality dependence of \mpt is present.

\subsection{\label{rcpsec}$R_{CP}$ in the forward directions}

\begin{figure*}[t]
\vskip -0.35cm
\begin{center}
\includegraphics[scale=0.475]{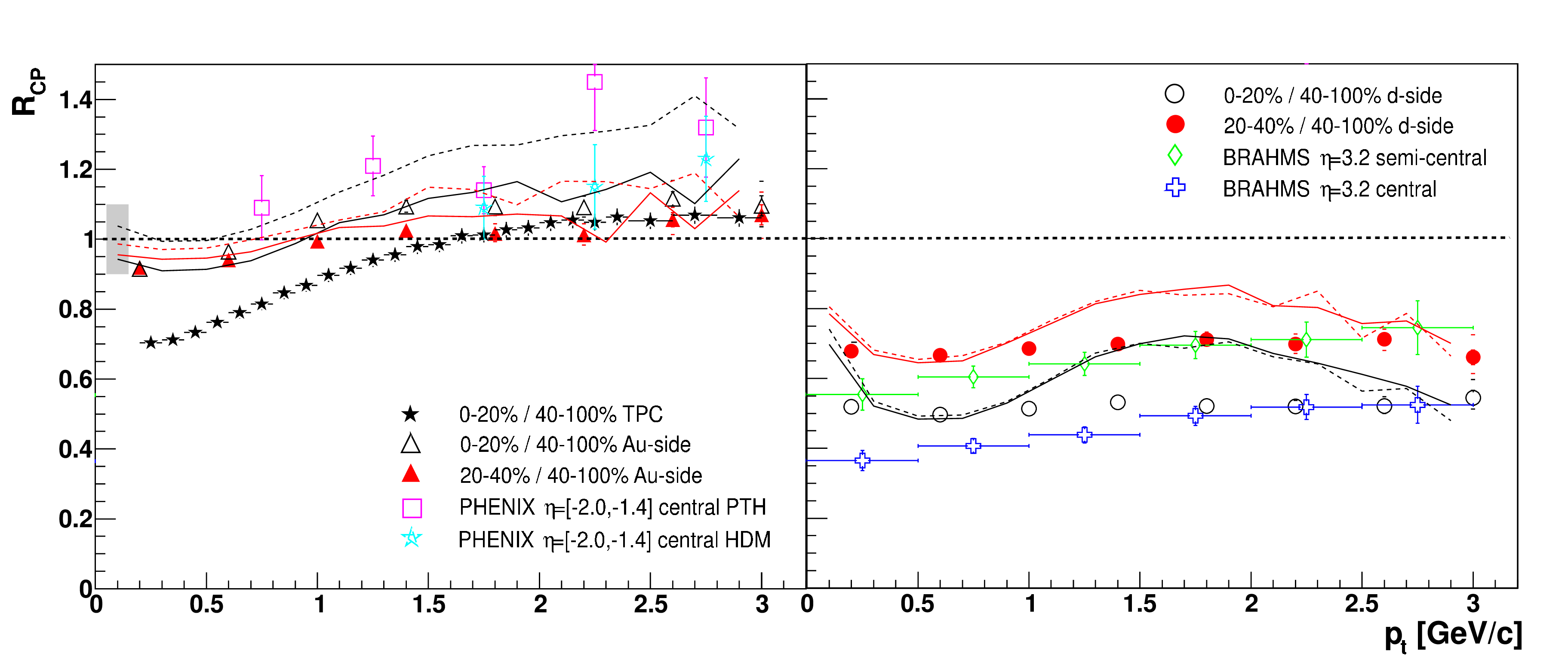}
\end{center}
\vskip -0.35cm
\caption{\label{fig2} (Color online) \rcp at $\eta \approx$  $-$3.1 (Au-side; triangles) and at $\eta \approx$  3.1 (d-side; circles) for 20\% and 20-40\% central d+Au events. The grey band represents the estimate of the systematic error \cite{thesis}. Midrapidity charged hadron \rcp (0-20\%) ($|\eta|<$ 0.5) is shown as stars \cite{johan}. Measurements from BRAHMS at $\eta \approx$  3.2 on the d-side \cite{brahms_dAu_rcp} and from PHENIX in the region of $-$2 $< \eta <$  $-$1.4 on the Au-side \cite{phenix_rcp} are also overlayed. The solid lines represent HIJING simulations \cite{hijing} without shadowing, the dashed lines with shadowing.}
\end{figure*}

\begin{figure*}[t]
\begin{center}
\includegraphics[scale=0.875]{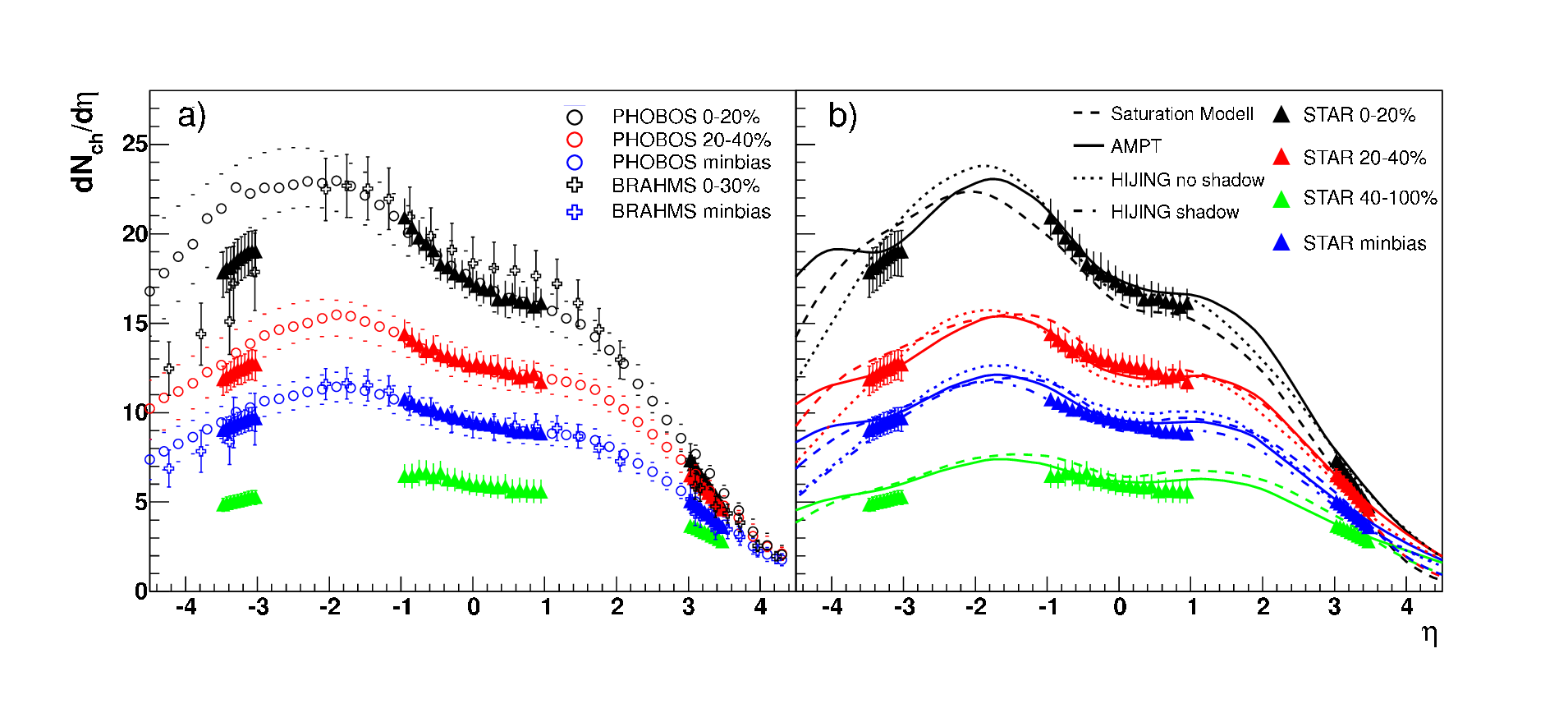}
\end{center}
\vskip -0.85cm
\caption{\label{fig3} (Color online) Charged hadron pseudorapidity distribution per event in the TPC and FTPC acceptance for 0-20\%, 20-40\%, 40-100\% central and minimum bias d+Au events (triangles). The error bars include both statistical and systematic error. Also overlayed are measurements from BRAHMS (crosses) \cite{brahms_dAu} and PHOBOS (circles) \cite{phob_dAu_mb, phob_dAu_qm04}. In addition predictions from HIJING \cite{hijing}, AMPT \cite{ua5} and the saturation model \cite{saturation_dAu} are also plotted compared with STAR measurements.}
\end{figure*}

Another variable of interest is the ratio of central to peripheral inclusive d+Au spectra

\begin{equation}
\label{eq:Rcp}
R_{CP}(p_t)=\frac{(d^2N/dp_t d\eta/\langle N_{bin}\rangle)|_{central}}{(d^2N/dp_t d\eta/\langle N_{bin}\rangle)|_{periph}}\,,
\end{equation}
\\
where $d^2N/dp_{t}d\eta$ is the differential yield per event and  $\langle N_{bin} \rangle\,$ the mean number of binary collisions for the corresponding centrality class, calculated using a Monte Carlo Glauber model \cite{star_dAu} (see Table \ref{tab1}). For the \rcp measurements the transverse momentum range could be expanded to 3 GeV/c (instead of 1 GeV/c for the transverse momentum measurements in section \ref{mptsec}) due to the centrality independence of the momentum resolution. Therefore, the effect of momentum resolution on the transverse momentum spectra cancels out in the \rcp ratio.

\begin{table}[htdp]
\begin{center}
\begin{tabular*}{0.45\textwidth}{@{\extracolsep{\fill}} c|c|c}
{\bf Centrality class} & \npart & \nbin \\
\hline
0-20\% & 15.69 $\pm$ 1.19 & 15.07 $\pm$ 1.29 \\
20-40\% & 11.17 $\pm$ 1.11 & 10.61 $\pm$ 0.80 \\
40-100\% & 5.14 $\pm$ 0.44 & 4.21 $\pm$ 0.49 \\
0-100\% &  8.31 $\pm$  0.37 & 7.51 $\pm$ 0.39 \\
\hline
\end{tabular*}
\end{center}
\caption{\npart\  and \nbin\  for various centrality classes.}
\label{tab1}
\end{table}

Comparing forward to backward rapidities, it can be seen in Fig.\ \ref{fig2} that \rcp is increasing with $p_{t}$ for $p_{t}<$ 3 GeV/c on the Au-side. Also \rcp is larger on the Au-side, which indicates that the Cronin effect is more pronounced on the Au-side of a d+Au collision. Strong centrality dependence of \rcp on the d-side of the collisions is another interesting feature seen in Fig.\ \ref{fig2}. This observation was also reported by the BRAHMS collaboration \cite{brahms_dAu_rcp}. However, no significant centrality dependence is observed on the Au-side of a d+Au collision. The \rcp measurements from the FTPCs are in good overall agreement with measurements from BRAHMS on the d-side \cite{brahms_dAu_rcp} for $p_t >$ 1 GeV/c and in agreement with PHENIX on the Au-side \cite{phenix_rcp} of a d+Au collision (see Fig.\ \ref{fig2}). The discrepancy between the BRAHMS and STAR \rcp measurements on the d-side at $\eta \approx$ 3 for $p_t<$ 1 GeV/c can not be completely resolved at this time. The discrepancy can be partially attributed to the different centrality classes used by the two experiments to calculate the \rcpno. BRAHMS uses 60-80\% as the most peripheral bin whereas STAR uses the 40-100\% centrality class.  Also, the BRAHMS centrality selection is biased towards peripheral collisions in forward rapidities as discussed in section \ref{dndetasec}. Furthermore, different low-$p_{t}$ cut-offs may affect the low $p_t$ measurements, where the difference between BRAHMS and STAR is most prominent.

The suppression of \rcp (and \rdau \footnote{$R_{AB}(p_t)=(d^2N/dp_t d\eta)/(T_{AB}  d^2\sigma^{pp}/dp_t d\eta)$, where $d^2N/dp_t d\eta$ is the differential yield per event in the nuclear collision A+B and $T_{AB}=\langle N_{bin}\rangle/\sigma^{pp}_{inel}$ describes the nuclear geometry with respect to the p+p reference measurements.}) at higher rapidities on the d-side is in qualitative agreement with predictions of the saturation model \cite{saturation_dAu}. Models based on pQCD which incorporate initial-state parton scattering and energy loss can also describe the behaviour of \rcp at higher rapidities [8, 9]. Furthermore, in the framework of parton recombination in the final state, \rcp at forward rapidities can be described as well \cite{recombination}.

\subsection{\label{dndetasec}Charged particle density asymmetry}

In Fig.\ \ref{fig3}a the pseudorapidity distribution $dN_{ch}/d\eta$ of charged hadrons per event in the TPC and FTPC acceptance is shown for minimum bias and for the 0-20\%, 20-40\%, and 40-100\% most central events. For comparison, measurements of the pseudorapidity distribution from BRAHMS \cite{brahms_dAu} and PHOBOS \cite{phob_dAu_mb, phob_dAu_qm04} are also plotted. The measured $dN_{ch}/d\eta$ distributions for minimum bias d+Au events are in good agreement for all three experiments. However, with increasing centrality, a significant difference in the particle density at negative pseudorapidity values $\eta <$  $-$3 between STAR and PHOBOS is visible. On the other hand, the measurements in the mid-pseudorapidity region are in good agreement. When comparing central events, for the $\eta <$  $-$3 region the BRAHMS $dN_{ch}/d\eta$ distribution is lower than the STAR measurements; at mid-rapidity it is higher. A possible explanation could be the different methods used for centrality selection. Centrality selection for the STAR-TPC was done via the \nch multiplicity in the FTPC and vice versa, to avoid autocorrelations caused by fluctuations in  the measured multiplicity. Simulation studies show that with a pseudorapidity gap of 2 units between the detectors this method is insensitive to autocorrelations \cite{thesis}. Use of the FTPC \nch multiplicity on the Au-side instead leads to a visibly higher particle density in the  $dN_{ch}/d\eta$ distribution for $\eta <$ $-$3, causing a significant bias in the centrality definition \cite{thesis}. This observation explains the higher particle density measured by PHOBOS for the Au-side of a d+Au collision, because their centrality was determined via the multiplicity in the pseudorapidity region of $-$4 $<  \eta <$  $-$3.5. For BRAHMS the enhancement in the particle density at midrapidity could be due to the fact that the multiplicity in the central region $|\eta| <$ 2.2 was used to define centrality. However, within the systematic errors the results of all three experiments are consistent with each other. 

In addition, the measured pseudorapidity distributions were compared with model predictions. Calculations based on gluon saturation in the Color Glass Condensate \cite{saturation_dAu} as well as results of HIJING \cite{hijing} and a Multi-Phase Transport Model (AMPT) \cite{ampt_dAu} are shown in  Fig.\ \ref{fig3}b. All model calculations are in good overall agreement with the measured $dN_{ch}/d\eta$ distributions for different centrality classes. In particular, the models are able to reproduce the increasing asymmetry of charged particle densities with increasing centrality.


\section{\label{yieldsec}Collision asymmetry and yield suppression}

\begin{figure}[t]
\begin{center}
\vskip -0.35cm
\includegraphics[scale=0.45]{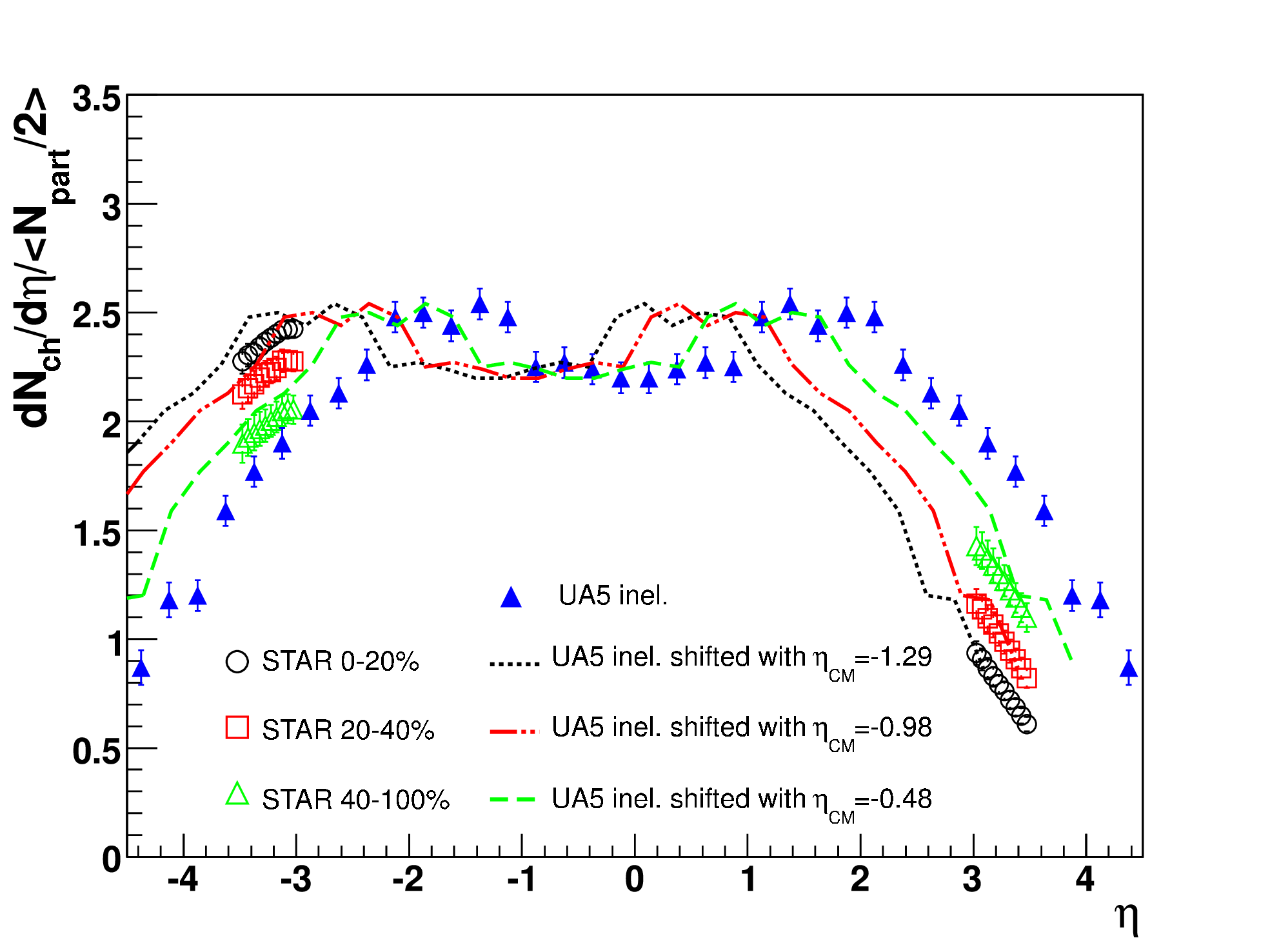}
\end{center}
\vskip -0.25cm
\caption{\label{fig4} (Color online) Pseudorapidity distribution of charged hadrons in the FTPC acceptance for 0-20\%, 20-40\%, 40-100\% central d+Au events scaled with Npart (open triangles). p+p measurements \cite{ua5} unshifted (triangles) and shifted by $\eta_{CM}$ (lines) are overlaid (for further details see text).}
\end{figure}

In the following section the centrality dependence of \rcp at high rapidities $|\eta| \approx$  3.1 will be discussed in connection with the observed increasing asymmetry of the produced particle density in d+Au collisions at transverse momenta $p_{t} <$ 3 GeV/c. In Fig.\ \ref{fig2}, the \rcp from HIJING simulations for different centralities with and without shadowing is shown. It is evident from Fig.\ \ref{fig2} that HIJING reproduces the overall behaviour of \rcp at $|\eta| \approx$ 3.1. In addition, it can be concluded that the influence of shadowing for $p_{t}>$ 1 GeV/c only affects the measurements on the Au-side. Since the $p_{t}$ spectrum on the d-side is more or less independent of centrality -- except for an overall scale -- (see Fig.\ \ref{fig1}a and b) and comparable with p+p collisions, the suppression of \rcp on the d-side (see Fig.\ \ref{fig2}) could be due to the asymmetry in particle production in d+Au collisions with respect to the symmetric p+p collisions (see Fig.\ \ref{fig4}). To take the asymmetry in d+Au collisions into account, a new variable $\eta_{CM}$ is introduced, which is defined as the weighted mean of the $dN_{ch}/d\eta$ distribution for each centrality class ($\eta_{CM}=\eta_{CM}(N_{part})$). $\eta_{CM}$ was extracted from the published PHOBOS results \cite{phob_dAu_qm04} and should represent the shift of the center-of-mass in the asymmetric d+Au collisions with respect to the nucleon-nucleon center-of-mass system (see Table \ref{tab2}). Even though the PHOBOS data are biased towards higher multiplicity on the Au-side as discussed in section \ref{dndetasec} they were used to determine $\eta_{CM}$ to maintain a model independent approach (also they are the only measurements available covering the full $\eta$ range). The $dN_{ch}/d\eta$ distribution for inelastic p+p collisions \cite{brahms_dAu} in this new reference system - obtained by shifting with $\eta_{CM}$ - are shown in Fig.\ \ref{fig4}. One observes that the d+Au $dN_{ch}/d\eta$ distributions normalized with $\langle N_{part}/2 \rangle\,$ (see Table \ref{tab1}) at high rapidities are consistent with the shifted p+p values. Also, the centrality dependence can be qualitatively explained. Therefore, $\eta_{CM}$ seems to be an appropriate variable to describe the asymmetry in particle production in d+Au collisions assuming that this asymmetry is caused by the nuclear stopping of the deuteron while traversing through the gold nucleus. Similar approaches to describe the pseudorapidity distributions in d+Au can be found in \cite{thesis,bari,steinberg}.

\begin{table}[htdp]
\begin{center}
\begin{tabular*}{0.25\textwidth}{@{\extracolsep{\fill}} c|c}
{\bf Centrality class} & $\eta_{CM}$ \\
\hline
0-20\% & -1.29 \\
20-40\% & -0.98 \\
40-60\% & -0.68 \\
60-80\% & -0.37 \\
80-100\% & -0.14 \\
40-100\% & -0.48 \\
0-100\% & -0.92 \\
\hline
\end{tabular*}
\end{center}
\caption{$\eta_{CM}$ defined as the weighted mean of the PHOBOS $dN_{ch/}d\eta$ distribution for various centrality classes \cite{phob_dAu_qm04}.}
\label{tab2}
\end{table}

In that representation, the suppression of the particle density on the d-side and the enhancement on the Au-side in asymmetric d+Au collisons with respect to the symmetric p+p collisions (see Fig.\ \ref{fig4}) can be expressed by defining \sdau as the ratio of the p+p reference $dN_{ch}^{pp}/d\eta$ distribution shifted with $\eta_{CM}$ ($dN_{ch}^{pp}/d\eta|_{\eta-\eta_{CM}}$) and the unshifted $dN_{ch}^{pp}/d\eta$ distribution:
\begin{equation}
\label{eq:scale}
S_{dAu}(\eta,\eta_{CM})=\frac{\left. dN_{ch}^{pp}/d\eta \right |_{\eta-\eta_{CM}}}{\left. dN_{ch}^{pp}/d\eta \right |_{\eta}} (\eta,\eta_{CM})\,.
\end{equation}
Our ansatz is then that the suppression and enhancement of \rdau (and \rcp ) is mainly caused by this geometric asymmetry. It would follow from this ansatz that the difference in the observed centrality dependence of \rdau should be accounted for by simply scaling with \sdau (Eq.\ \ref{eq:scale}). The BRAHMS \rdau measurements \cite{brahms_dAu_rcp} at different pseudorapidities ($\eta$=0, 1, 2.2 and 3.2) on the d-side are then consistent with a universal behaviour, reaching binary scaling at $p_{t}>$ 3 GeV without a significant Cronin enhancement at intermediate $p_{t}$.

\begin{figure}[t]
\begin{center}
\vskip -0.35cm
\includegraphics[scale=0.45]{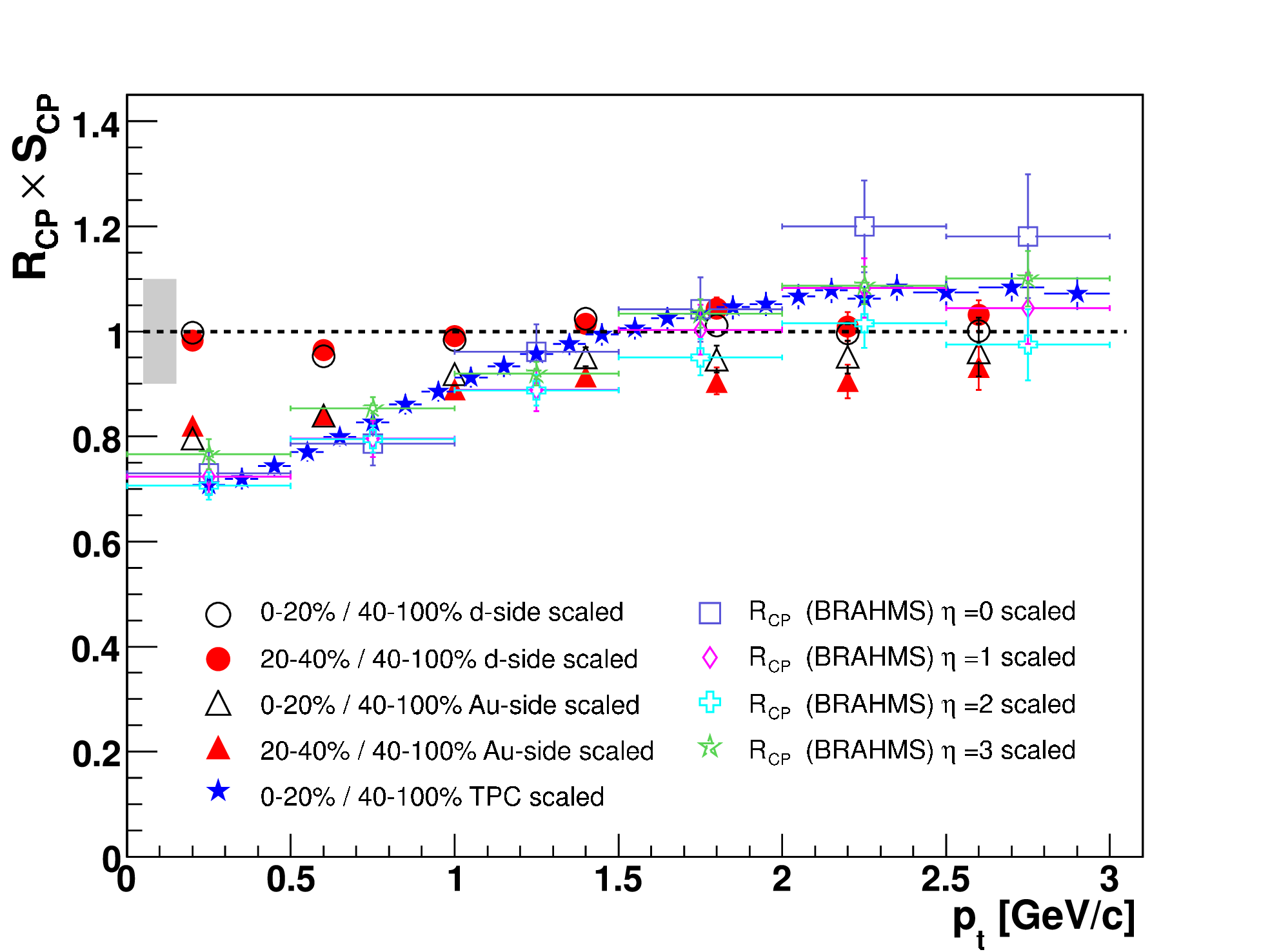}
\end{center}
\vskip -0.25cm
\caption{\label{fig5}(Color online)  \rcp of charged hadrons at $\eta \approx$  $-$3.1 (Au-side) and at $\eta \approx$  3.1 (d-side) as function of pt for different centrality classes scaled with \scp (Eq.\ \ref{eq:scaledau}). Midrapidity scaled charged hadron \rcp (0-20\%) ($|\eta|<$ 0.5) is shown as stars \cite{johan}. Also plotted are scaled BRAHMS \rcp measurements (0-20\%/60-80\%) \cite{brahms_dAu_rcp} in different $\eta$ intervalls.}
\end{figure}

A similar procedure can be applied to describe \rcp measurements for different centrality classes and pseudorapidities on the d- and Au-side of a d+Au collision. One has to modify \sdau (Eq. 2) to take the asymmetry in particle production - still visible in peripheral d+Au collisions (see Fig.\ \ref{fig4}) - into account. This is realized by taking the peripheral d+Au $dN_{ch}/d\eta$ distribution as reference and shifting the denominator in Eq.\ \ref{eq:scale} according to the peripheral $\eta_{CM}$ value:
\begin{equation}
\label{eq:scaledau}
S_{CP}(\eta,\eta_{CM})=\frac{\left. dN_{ch}^{pp}/d\eta \right |_{\eta-\eta_{CM}}}{\left. dN_{ch}^{pp}/d\eta \right |_{\eta-\eta_{CM,periph.}}} (\eta,\eta_{CM})\,.
\end{equation}
\rcp measurements from the FTPCs at $|\eta| \approx$ 3.1 for different centrality classes scaled with \scp are shown in Fig.\ \ref{fig5}. These results suggest that the observed enhancement of \rcp on the Au-side for different centrality classes can be explained by an enhancement of particle production caused by the nuclear stopping of the deuteron and described by a shift of the center-of-mass with respect to the most peripheral d+Au collisions. The suppression of \rcp on the d-side is explained analogously. Universal scaling behaviour of \rcp  scaled with \scp at $|\eta| \approx$ 3.1 is visible for $p_t>$ 1 GeV/c  in the FTPC measurements (see Fig.\ \ref{fig5}). In addition scaled BRAHMS \rcp measurements for the top 20\% central d+Au events at different pseudorapidities are also overlayed in Fig.\ \ref{fig5}. Universal scaling behaviour of scaled \rcp is seen in both STAR and BRAHMS measurements reaching binary scaling for $p_t  >$ 2 GeV/c.  The deviation from the supposed scaling for the BRAHMS \rcp measurements at midrapidity (and also the discrepancy to the STAR midrapidty measurements) might be explained by the different centrality definitions, where a bias to higher particle multiplicities might be present in the BRAHMS data as discussed in section \ref{dndetasec}. It should be noted that systematic errors in the scaling due to uncertainties in the determination of $\eta_{CM}$ were not estimated in this analysis. One would expect an overall normalization uncertainty, but the main feature of scaling should be preserved. The simple data driven picture outlined in this paper was meant to point out the importance of the collision geometry in d+Au collisions.

\section{Conclusions}

Centrality dependence of  the $dN/d\eta$ distributions in $\sqrt{s_{NN}}$=200 GeV d+Au collisions is presented. Within errors, $dN/d\eta$ cannot discriminate between different model calculations. However, our data show an increase of \mpt as a function of centrality on the Au-side, whereas on the d-side no centrality dependence is visible. This result together with the suppression of \rcp on the d-side and its enhancement on the Au-side relative to mid-rapidity cannot be described consistently by current model calculations \cite{accardi04}. A similar study of comparing particle production at high $p_t$ in d- and Au-side at mid-rapidity \cite{rapasym} has ruled out models based on incoherent initial multiple partonic scattering and independent fragmentation. It also showed that models based on nuclear shadowing incorporating extremes of gluon shadowing at low $x$ can not account for the difference in particle production in d- and Au-side at high $p_t$ at mid-rapidity.

This paper demonstrated, that in a simple stopping picture the main features of the pseudorapidity and centrality dependence of \rcp (and \rdauno) at higher rapidities can be explained by the suppression (enhancement) of particle yields in d+Au relative to p+p collisions and peripheral d+Au collisions. Simulation studies show a small effect of shadowing in HIJING on \rcpno, especially on the d-side, supporting this geometric picture. This result is also confirmed by measurements of $R_{pPb}$ by NA49 \cite{na49rcp} at an order of magnitude lower energies, where shadowing and gluon saturation are expected to be small, which show qualitatively the same characteristic  centrality and rapidity ($x_F$) dependence and are consistent with the stopping picture.

Taking the asymmetry in particle production -- characterized by a shift of the center-of-mass in the asymmetric d+Au collisions with respect to the nucleon-nucleon or peripheral d+Au center-of-mass system -- into account, \rcp (and \rdauno) show a universal scaling behaviour independent of centrality and pseudorapidity in d+Au collisions at RHIC energies.
On the other hand, the success of the CGC saturation approach \cite{saturation_dau_new} and stopping picture in quantitatively describing BRAHMS  \rcp (and \rdauno) measurements could be interpreted as a link between saturation as the possible origin of nuclear stopping. In that case deviations from the observed empirical scaling in asymmetric collision systems at different energies can help to quantify the onset of saturation effects in heavy-ion collisions.\\  

We thank the RHIC Operations Group and RCF at BNL, and the
NERSC Center at LBNL for their support. This work was supported
in part by the HENP Divisions of the Office of Science of the U.S.
DOE; the U.S. NSF; the BMBF of Germany; IN2P3, RA, RPL, and
EMN of France; EPSRC of the United Kingdom; FAPESP of Brazil;
the Russian Ministry of Science and Technology; the Ministry of
Education and the NNSFC of China; IRP and GA of the Czech Republic,
FOM of the Netherlands, DAE, DST, and CSIR of the Government
of India; Swiss NSF; the Polish State Committee for Scientific 
Research; STAA of Slovakia, and the Korea Sci.\ \& Eng.\ Foundation.\\

\bibliography{ref}

\end{document}